\documentclass[a4paper,11pt]{article}
\usepackage{latexsym}
\usepackage{cite}
\usepackage{epsfig}
\usepackage{amssymb}
\usepackage{latexsym}
\usepackage{epsfig, subfigure, subfloat, graphicx, float}
\usepackage{mathrsfs}
\usepackage{amsmath}
\usepackage{color}
\usepackage[usenames]{xcolor}
\graphicspath{{Image/}}
\author{H. Mohseni Sadjadi\footnote{mohsenisad@ut.ac.ir}  and M. Khodaei\footnote{mahdi.khodaei@ut.ac.ir}
\\ {\small Department of Physics, University of Tehran,}
\\ {\small P. O. B. 14395-547, Tehran 14399-55961, Iran}}
\title{Regular scalar field around reflecting stars and black holes, and reflecting star polarization}
\begin{document}
\maketitle
\begin{abstract}
The existence of a regular non-trivial scalar field in the background of an asymptotically flat static reflecting star is discussed. The scalar field is assumed to be conformally coupled to the outside matter. The induced scalar charge is determined and the required conditions to have regular field are obtained. The results are compared with the case of a black hole. Conditions to have regular non-trivial scalar field out of the black hole horizon are investigated. We show that in the presence of a scalar source, in contrast to the black holes, the reflecting star becomes polarized.
\end{abstract}

\section{Introduction}

The only scalar field in the standard model of particles is the Higgs boson\cite{H,H1,H2}. But scalar fields are well motivated in the beyond standard models of particles and cosmology, especially in describing inflation, dark energy and dark matter \cite{qu}. The study of scalar fields in space-time with non-trivial geometry or topology, or with different boundary conditions, has been the subject of many studies. One of these interesting backgrounds is the black hole, whose horizon absorbs irreversibly radiation and matter. A question that naturally arises is whether there can be non-trivial (non-constant) scalar fields outside the horizon of a black hole \cite{radu}. This may be studied by solving the scalar field equation. The absence of such a field is a confirmation of the no-hair conjecture. This conjecture states that the black hole solutions of Einstein-Maxwell equations in general relativity are characterized by three parameters: mass, electric charge, and angular momentum. In this view black-holes support only gravitation and electric fields (fields associated with the Gauss law) outside their horizons \cite{nohair}. In \cite{scalarmassless}, it was shown that a regular massless scalar field cannot exist in the background of a static asymptotically flat black hole. In \cite{scalarmass}, without referring to the Einstein equations, it was shown that canonical minimally coupled scalar fields, $\phi$, with potentials satisfying $\phi V_{,\phi}>0$, do not exist in the exterior of a stationary black hole background. Many models have been proposed to examine the conjecture of \cite{scalarmass} and to find counterexamples by changing the basic assumptions upon which this conjecture was initially based. This can be accomplished by modifying the general theory of gravity, or by using non-canonical or non-minimally coupled scalar fields and so on (see \cite{radu} and references therein). For example, by coupling the scalar field to the Ricci scalar, through a term proportional to $R \phi^2$, one can find a black hole solution (with a secondary hair, in the sense that the scalar field is not attributed to a new charge), although the scalar field is not regular at the horizon \cite{beck2,beck21,beck22}.

Apart from a black hole, the endpoint of a gravitational collapse may be a compact star without any singularity and event horizon \cite{Mazur}. This may occur in some quantum-gravity models, where quantum effects may prevent the formation of classical black hole horizons. In \cite{Ch}, a gravastar model (a very compact star without an event horizon or singularity with a radius close to the Schwarzschild radius) was introduced, with the same gravitational redshift in the electromagnetic emission as a black hole.
As was studied in \cite{Barc,Card}, an ultra-compact black horizonless star surface shows reflective properties. This inspires the intriguing idea to replace the horizon by a reflecting surface.

To see whether the no existence theorem for canonical scalar fields, is an exclusive characteristic of black-holes, the same problem was considered in the background of a compact object with a reflecting (repulsing) surface (instead of the absorbing event horizon) \cite{hod1}. The domain of the scalar field is bounded by a boundary where the field vanishes and the spatial infinity where the field tends to zero. A physical compact object whose on the surface the external field vanishes is dubbed "reflecting star" in \cite{hod1}. We use this terminology in this paper. \cite{hod1}, and subsequent sequel papers \cite{hod2,hod21,hod22,hod23}, \cite{sark}, and \cite{peng,peng1,peng2,peng3}, show that the reflecting star does not support canonical massive scalar field outside its surface. In \cite{peng4}, it is shown that even by considering the matter field's back reaction, there is not scalar hair for a large reflecting star. This is similar to the no scalar hair for large black holes \cite{L1,L2}.

So, interestingly it seems that this space-time satisfies the no-scalar hair conjecture similar to an asymptotically flat black hole. We note that, in this issue, there is a crucial difference between black holes and reflecting stars. In the black hole background, scalar and electric fields satisfy completely different equations of motion, i.e. Maxwell and Klein-Gordon equations. So although we have not a regular non-trivial scalar field in the background of a black hole, we may have a regular electric field on its horizon and outside it. As we will see in the third section (3.1), this is not true for reflecting stars, i.e. there is no scalar field and no electric field outside the reflecting star. This may be easily verified by solving the Maxwell equations by complying with the imposed boundary condition. This situation is somehow similar to a grounded black hole, pointed out in \cite{Mayo}.

 Matter outside the black hole, by interacting with the scalar field, may change its profile and permits to have a no trivial regular solution, which was not allowed before (in this situation the solutions are no more scalar-vacuum solutions). Screening dark energy models \cite{screen,screen01,screen02,screen03} may be very inspiring in this issue. In this context, the interaction of the scalar field and matter, through a conformal coupling, allows a nontrivial regular scalar field profile, provided that some constraints be satisfied \cite{screen1,screen11}. Deriving analytical solution for a conformally coupled scalar field in the background of a black hole, even by ignoring the back-reaction on the metric, is not generally possible and must be accomplished numerically. In this situation, employing a Dirac-delta like test charge, which enables us to obtain analytical solutions may be very useful. Considering the charge as a source of the scalar or electric fields outside the horizon has been studied extensively for deriving forces on particles outside the horizon \cite{selforce,selforce1,selforce2,selforce3,selforce4}, and also examining the no-hair conjecture \cite{testnohair,testnohair1,testnohair2}.  In this context, by only solving the field equation and without referring to the full Einstein equations and back-reaction effects, one may study the black hole's ability to admit charges (free or induced) and to acquire hair\cite{Mayo}.

In this paper, motivated by the above discussions, we aim to study the existence of a regular nontrivial massless scalar field in the background of an asymptotically flat reflecting star and compare the results with the black hole case. Inspired by screening models, we perform our study in the presence of matter which is conformally coupled to the scalar field. In the second section, we show how this conformal coupling may alter the non-existence theorem pointed out in \cite{scalarmass} and \cite{hod1}.  To explain the subject through analytical solutions, we consider a spherically symmetric test charge and obtain regular scalar field analytical solutions and discuss the reflecting star polarization and determine the induced scalar charge. We elucidate our results by choosing some specific conformal factors and metrics.

 We use units $\hbar=c=G=1$.

\section{Non-trivial conformally coupled scalar fields and non existence theorem}
We consider the Lagrangian \cite{screen}
\begin{equation}\label{2}
\mathcal{L}=\frac{1}{16\pi}R-\frac{1}{2}\partial_\mu \phi \partial^\mu \phi+\mathcal{L}_m(\psi_m^{(i)},g_{\mu\nu}^{(i)}).
\end{equation}
The matter field $\psi_m^{(i)}$, described by the Lagrangian $\mathcal{L}_m$, is
coupled to the metric $g_{\mu\nu}^{(i)}$  which in terms of the Einstein
frame metric $g_{\mu \nu}$ is given by
\begin{equation}\label{3}
 g_{\mu\nu}^{(i)}=A^2(\phi)g_{\mu \nu}.
\end{equation}
So the massless scalar field is interacting with the matter component $\psi_m^{(i)}$, through a conformal coupling. Such a coupling has been vastly used to study the scalar field dark energy models \cite{trod}, and also screening in dense regions \cite{kho}.
The equation of motion of the scalar field is obtained in \cite{screen,kho} as
\begin{equation}\label{4}
\Box \phi=-(lnA)_{,\phi}T_{(m)},
\end{equation}
where the trace of matter energy momentum tensor is $T_{(m)}=T^{\mu \nu}_{(m)}g_{\mu \nu}$. It is convenient to define $\rho=-A^{-1}T_{(m)}$ which is the conserved non relativistic matter density \cite{screen}. So we have (for a full derivation see \cite{screen01} and its eq.(20))
\begin{equation}\label{5}
\Box \phi=A_{,\phi}\rho,
\end{equation}
where $\rho$ is the density corresponding to the pressureless matter$\psi_m^{(i)}$. Note that the conformal coupling in (\ref{2}) is considered only for the matter component $\psi_m^{(i)}$, i.e. $\phi$ interacts only with  $\psi_m^{(i)}$, and not with the other components. Such different couplings to different components have been studied in the literature \cite{trod,sar}. In addition as our study is restricted to the outside of the reflecting surface, then the inside density does not appear in the local equation (\ref{5}).

If $A_{,\phi}= cte$ , (\ref{5}) reduces to the field equation in the presence of the source $\rho$ , and for $A_{,\phi}\propto \phi$, (\ref{5}) describes a scalar field with an squared effective mass term proportional to $\rho$.  In the absence of $A_{,\phi}\rho$, we are left with $\Box \phi=0$. This equation has not regular non-trivial (by trivial we mean constant) static solution in the background of a black hole. Based on no-hair conjecture, stationary black holes do not admit scalar hair in the minimal case in the general theory of relativity \cite{scalarmassless}. The same occurs for reflecting stars:  $\Box \phi=0$ has not static nontrivial solution satisfying $\phi(r\to \infty)=0$ and $\phi(r_s)=0$ where $r=r_s$ is the star surface \cite{hod1}.  In our study, if (\ref{1}) has an horizon located at $r_h$, we take $r_s>r_h$. In the following, we aim to study whether the source term $A_{,\phi}\rho$, arisen from the conformal coupling, alerts these nonexistence theorems in the background of a stationary reflecting star.

By using (\ref{5}), one obtains
\begin{equation}\label{6}
\int_{\partial{\mathcal{V}}} \phi\nabla^\mu \phi \sqrt{h}  n_\mu d^3\sigma -\int_{\mathcal{V}} \nabla_\mu \phi \nabla^\mu \phi \sqrt{-g} d^4x=\int_{\mathcal{V}} \phi A_{,\phi }\rho \sqrt{-g} d^4 x,
\end{equation}
where the region under study, $\mathcal{V}$, is bounded by the reflecting surface and spatial infinity (and in time, by two partial Cauchy surfaces \cite{hw} where their contributions cancel each other \cite{radu}). $h_{\mu \nu}$ is the induced metric on the boundary.  The contribution of the reflecting surface, where the scalar field is zero, vanishes and the field is assumed to fall off sufficiently fast at infinity. The same is true for scalar field in the asymptotically flat  black hole background where the reflecting surface boundary is replaced by a Killing horizon \cite{scalarmass}. So we have
\begin{equation}\label{7}
-\int_{\mathcal{V}} \nabla_\mu \phi \nabla^\mu \phi \sqrt{-g} d^4x=\int_{\mathcal{V}} \phi A_{,\phi }\rho \sqrt{-g} d^4 x.
\end{equation}
In the absence of the source term, the right-hand side would be zero implying $\phi=cte$ which for reflecting stars, where $\phi(r_s)=0$, gives $\phi(r>r_s)=0$.
This in agreement with \cite{hod1}, where this theorem was proved in another way. It is clear that in the presence of the source term in the right-hand side of (\ref{7}), alters this statement. The left hand side of (\ref{7}) is negative, hence a necessary condition to have nontrivial scalar field is $\int_{\mathcal{V}} \phi A_{,\phi }\rho \sqrt{-g} d^4x <0$. For example for a positive $\rho$, and $A_{,\phi}=\frac{1}{M}\phi$, we must have $M<0$.

Now let us  multiply both sides of (\ref{5}) by $A_{,\phi}$, to obtain
\begin{equation}\label{8}
\int_{\mathcal{V}} \left(A_{,\phi}\nabla^\mu \nabla_\mu \phi-A_{,\phi}^2\rho\right)\sqrt{-g}d^4x=0.
\end{equation}
After some calculation we find
\begin{equation}\label{9}
\int_{\partial{\mathcal{V}}} \left(A_{,\phi}\nabla_\mu \phi\right) n^\mu \sqrt{h}d^3x-\int_{\mathcal{V}} \left(A_{,\phi \phi}\nabla^\mu \phi \nabla_\mu \phi +A_{,\phi}^2\rho\right)\sqrt{-g}d^4x=0
\end{equation}
The region and its boundary are the ones mentioned above.  If $A_{,\phi}=0$ on $r=r_s$ (e.g. for a power law $A$, where $A\propto \phi^n,\,\,n>1$), and for sufficiently fast decaying scalar field at infinity, we obtain
\begin{equation}\label{10}
\int_{\mathcal{V}} \left(A_{,\phi \phi}\nabla^\mu \phi \nabla_\mu \phi +A_{,\phi}^2\rho\right)\sqrt{-g}d^4x=0.
\end{equation}
 Hence if everywhere $\rho A_{,\phi \phi}>0 $, only a trivial $\phi$ satisfying  $A_{,\phi}=0$ is allowed. Conversely if only a trivial $\phi$ is allowed, then no-hair conjecture requires $A_{,\phi}=0$. This can also be verified in another way:  If there is no solution for $A_{,\phi} =0$ (e.g. $A(\phi)\propto \phi$ or $A(\phi)\propto exp(\beta\phi)$), there is no trivial solution for (\ref{5}). $\rho A_{,\phi \phi}$ is the effective mass squared corresponding to the field $\phi$:  when $\phi$ is linearly perturbed around the vacuum, by inserting $\phi=\phi_0+\delta\phi$ in (\ref{5}), we find
 \begin{equation}
 \Box \delta \phi= \rho A_{,\phi \phi}(\phi_0)\delta \phi.
 \end{equation}
 So we can evade no-hair theorem when this squared mass is negative.  The same argument was used in \cite{scalariz,scalariz1}, to explain scalarization. Additional couplings of the scalar field, permitting evading no existence theorem are also studied in \cite{bk1,bk2,bk3}.

 In the following, we try to solve (\ref{5}). As for a general $\rho$, obtaining an analytical solution for (\ref{5}) is not possible, we consider the matter density as a Dirac delta-like test charge. The convenient of this choice is that it provides the possibility to obtain an analytical solution for the scalar field. The study of test charges in the background of black holes was widely used in the literature \cite{testnohair}, to get a fundamental insight into the no-hair theorem and the capability of the black hole to support scalar fields and so on, without engaging in complicated equations which have not analytical solutions.

\section{Scalar test charge, the regular static scalar field and an induced scalar charge}
On the static spherically symmetric space-time
\begin{equation}\label{1}
ds^2=-e^{\mu(r)}dt^2+e^{\nu(r)}dr^2+r^2\left(d\theta^2+sin^2\theta d\phi^2\right),
\end{equation}
we consider a spherically symmetric scalar test charge (a shell with radius $R$). The static (so pressureless) density is given by \cite{wein}
\begin{equation}\label{11}
\rho_s=\frac{\lambda\delta(r-R)}{4\pi r^2 \sqrt{e^{\mu(r)} e^{\nu(r)}}}.
\end{equation}
With this source, in the static case, and with the metric (\ref{1}),(\ref{5}) becomes
\begin{equation}\label{12}
\frac{d}{dr}f(r)\frac{d}{dr}\phi=\lambda A_{,\phi}\delta(r-R),
\end{equation}
where
\begin{equation}\label{13}
f(r)=4\pi e^{-\frac{\nu(r)}{2}}e^{\frac{\mu(r)}{2}}r^2.
\end{equation}
In the absence of the coupling, the zeroes of $f(r)$ correspond to singularities of the solution. For example, solving   $\frac{d}{dr}f(r)\frac{d}{dr}\phi=0$, in a black hole background gives a singular scalar field at the horizon. This situation, as we will see, changes when the source term is considered.

General solution of (\ref{12}) is
\begin{equation}\label{14}
\phi = \left\{ \begin{array}{ll}
C_2F(r)+C_1& \textrm{if $r\geq R$}\\
C_4F(r)+C_3 & \textrm{if $r\leq R$},\\
\end{array} \right.
\end{equation}
where
\begin{equation}\label{15}
\int^r\frac{dr'}{f(r')}\equiv F(r).
\end{equation}
Asymptotically flatness requires $\lim_{r\to \infty}e^{\mu(r)}=1$ and $\lim_{r\to \infty}e^{\nu(r)}=1$. Therefore by counting superficial degree of (\ref{15}),  we obtain  $\lim_{r\to \infty}F(r)=0$. We demand also  $\lim_{r\to \infty}\phi(r)=0$, hence $C_1=0$. By integrating (\ref{12}), we obtain
\begin{equation}\label{16}
lim_{\epsilon \to 0}\int_{R-\epsilon}^{R+\epsilon} \left(\frac{d}{dr}f(r)\frac{d}{dr}\phi\right) dr=\lambda A_{,\phi}(R)
\end{equation}
which gives
\begin{equation}\label{r1}
f(r)\frac{d\phi}{dr}\Big|_{R+\epsilon}-f(r)\frac{d\phi}{dr}\Big|_{R-\epsilon}=\lambda A_{,\phi}(R).
\end{equation}
By using (\ref{14}) and (\ref{15}), (\ref{r1}) reduces to $C_2-C_4=\lambda A_{,\phi}(R)$.  Continuity of $\phi$ at $R$  necessitates  $C_2F(R)=C_4F(R)+C_3$. By applying the boundary condition  $\phi(r_s)=0$, where $r_s<R$, we obtain $C_3=-C_4F(r_s)$. Gathering all together, we obtain the final result for the reflecting star as
\begin{equation}\label{17}
\phi = \left\{ \begin{array}{ll}
\lambda A_{,\phi}(R)F(r)-\lambda\frac{F(R)}{F(r_s)}  A_{,\phi}(R)F(r)& \textrm{if $r\geq R$}\\
\lambda A_{,\phi}(R)F(R)-\lambda\frac{F(R)}{F(r_s)}  A_{,\phi}(R)F(r)& \textrm{if $r_s\leq r\leq R$}\\
\end{array} \right.
\end{equation}
 This look likes the potential of the original test charge shell, plus the potential of an induced spherical charge located at $r<r_s$ (because the only charge outside $r_s$ is the original charge $\lambda$)
 \begin{equation}\label{18}
 \lambda'=-\frac{F(R)}{F(r_s)}\lambda.
 \end{equation}
In other words $\phi$ is originated both from the source shell and an induced charge adopted by the reflecting star (so the reflecting star gets scalar hair). For $R=  r_s$, we have $\lambda'= -\lambda$, i.e. the sum of the scalar charges (free+induce) become zero, leading to $\phi(r)=0$ (no-hair) in agreement with \cite{hod1}. This shows that the star adopted an induced charge in the presence of the test external matter.

A necessary condition for existence of nontrivial solution is
\begin{equation}\label{19}
\phi(R)=\lambda A_{,\phi}(R)F(R)-\lambda \frac{F(R)}{F(r_s)} A_{,\phi}(R)F(R),
\end{equation}
which is derived by setting $r=R$ in (\ref{17}). Depending on the form of $A(\phi)$, (\ref{19}) restricts the parameters.

Without the boundary condition on $r=r_s$, the solution would be
\begin{equation}\label{20}
\phi = \left\{ \begin{array}{ll}
\lambda A_{,\phi}(R)F(r)+C_4F(r)& \textrm{if $r\geq R$}\\
\lambda A_{,\phi}(R)F(R)+C_4F(r)& \textrm{if $r\leq R$}\\
\end{array} \right.
\end{equation}
For a black hole with a horizon located at $r_h<R$, where $F(r_h)$ is singular, nonsingular solution is obtained by setting $C_4=0$
\begin{equation}\label{21}
\phi = \left\{ \begin{array}{ll}
\lambda A_{,\phi}(R)F(r)& \textrm{if $r\geq R$}\\
\lambda A_{,\phi}(R)F(R)& \textrm{if $r\leq R$}.\\
\end{array} \right.
\end{equation}
One important feature of this solution is that it is regular on the horizon. Note that there are not nontrivial regular solution in the absence of the source, i.e if we set $\lambda=0$ we obtain $\phi=0$ in agrement with \cite{scalarmassless}. Existence of non-trivial solutions requires that
\begin{equation}\label{c}
\phi(R)=\lambda A_{,\phi}(R)F(R),
\end{equation}
constraining the parameters of the model.

In the above study, the back-reaction of the test charge on the metric in solving the scalar field equation would give a higher-order term of $\lambda$ ($\mathcal{O}(\lambda^\alpha)$, with $\alpha>1$) which has been ignored in the small test charge limit analysis.

We conclude that in the presence of the external matter (\ref{11}), the nonexistence conjectures change: 1- reflecting star acquires an induced scalar charge 2- black hole adopts a regular scalar field in its background. To more elucidate these points, let us use some specific forms for the conformal factor $A$, and for the metric.

 \subsection{$A=M_0+\frac{\phi}{M}$}
As a first example we take the simplest case: $A=M_0+\frac{\phi}{M}$, where $M_0$ and $M$ are two real constants.  Based on our discussion after (\ref{10}), for $\lambda\neq 0$, only nontrivial solutions exist. The equation (\ref{5}) becomes
 \begin{equation}\label{22}
 M\nabla^2 \phi=\rho,
 \end{equation}
 which for the metric (\ref{1}), and the density (\ref{11}), reduces to
\begin{equation}\label{23}
M \frac{d}{dr}f(r)\frac{d}{dr}\phi=\lambda\delta(r-R).
\end{equation}
The solution is
 \begin{equation}\label{24}
\phi = \left\{ \begin{array}{ll}
\frac{ \lambda F(r)}{M}-\frac{\lambda F(R)}{MF(r_s)}F(r)& \textrm{ $r\geq R$}\\
\frac{\lambda F(R)}{M}-\frac{\lambda F(R)}{MF(r_s)}F(r)& \textrm{ $r_s\leq r\leq R$},\\
\end{array} \right.
\end{equation}
which may be rewritten as  $\phi=\phi_I+\phi_{II}$, where  $\phi_I$ is the potential due to the test charge,
\begin{equation}\label{25}
\phi_I = \left\{ \begin{array}{ll}
\frac{\lambda F(r)}{M},& \textrm{$r\geq R$}\\
\frac{\lambda F(R)}{M},& \textrm{$r_s\leq r\leq R$},\\
\end{array} \right.
\end{equation}
and the potential due to the induced charge $\lambda'=-\frac{F(R)}{F(r_s)}\lambda$ is
\begin{equation}\label{26}
\phi_{II}=\frac{\lambda'}{M}F(r), \,\,\ r\geq r_s.
\end{equation}
For the Schwarzschild metric
\begin{equation}\label{27}
ds^2=-\left(1-\frac{r_h}{r}\right)dt^2+\left(1-\frac{r_h}{r}\right)^{-1}dr^2+r^2d\Omega^2,
\end{equation}
we have
\begin{equation}\label{28}
\phi=\phi_I+\phi_{II} = \left\{ \begin{array}{ll}
\frac{\lambda}{4\pi M r_h}\ln \frac{r-r_h}{r}+ \frac{\lambda'}{4\pi M r_h}\ln \frac{r-r_h}{r},& \textrm{$r\geq R$}\\
\frac{\lambda}{4\pi Mr_h}\ln \frac{R-r_h}{R}+ \frac{\lambda'}{4\pi M r_h}\ln \frac{r-r_h}{r},& \textrm{$r_s\leq r\leq R$}\\
\end{array} \right.
\end{equation}
where
\begin{equation}\label{29}
\lambda'=-\frac{\ln \frac{R-r_h}{R}}{\ln \frac{r_s-r_h}{r_s}}\lambda.
\end{equation}
Note that $r_s>r_h$. As explained in the general case, (\ref{28}) shows that the we can have a regular nontrivial scalar field whose sources are the test scalar charge $\lambda$ and an induced scalar charge $\lambda'$ inside the reflecting star. For $R=r_s$, the reflecting star loses its hair.

Without the $r_s$ surface, the induced charge is removed and the solution reduces to
\begin{equation}\label{30}
\phi(r)= \left\{ \begin{array}{ll}
\frac{\lambda}{4\pi M r_h}\ln \frac{r-r_h}{r},& \textrm{$r\geq R$}\\
\frac{\lambda}{4\pi M r_h}\ln \frac{R-r_h}{R},& \textrm{$r\leq R$}\\
\end{array}\right.
\end{equation}
which is regular on the horizon, $r=r_h$. By putting $R=r_h$, we obtain
\begin{equation}\label{31}
\phi= \frac{\lambda}{4\pi M r_h}\ln \frac{r-r_h}{r},\,\,\,  r\geq r_h,
\end{equation}
 which is the same as the solution obtained for a test scalar field in \cite{Jan} with $\frac{\lambda}{M} \ll r_h$ (to be allowed to ignore the back-reaction). (\ref{31}), in contrast to (\ref{30}) is singular at $r=r_h$.

 Now let us see what would happen if we considered an electrostatic field in the background of the reflecting star?  In the usual black hole models, a black hole may support an electric charge whereas it has not scalar hair. Is this the same for the reflecting star? or the boundary condition forces $A_0$ to vanish too. To see this, let us consider an electrostatic field $A_0$ and a Schwarzschild like black hole space-time,  $e^{\frac{\mu(r)}{2}} e^{\frac{\nu(r)}{2}}=1$ . By considering the metric (\ref{27}), the corresponding equation, without an external source outside the horizon, is
 \begin{equation}\label{32}
\frac{d}{dr}4\pi r^2\frac{d}{dr}A_0=0
\end{equation}
The nontrivial solution vanishing at $r\ \to \infty$ is $A(r)=\frac{q}{4\pi r}$,  and the solution  which vanishes both at $r=r_s$ and $r\to \infty$ is $A_0=0$. So it seems that no-hair conjecture for the reflecting star, in contrast to black holes, includes also the electric charge. This situation is similar to a grounded black hole studied in \cite{Mayo}, i.e. a black hole whose horizon has zero potential and zero net electric charge. Note that this does not mean that the grounded black does not accept electric charges, rather it indicates that the sum of the charges (positive+negative) is zero. The ability to accept an electric charge by a grounded black hole can be investigated by inserting an electric test charge outside the horizon. To do so, as before, we consider a spherically symmetric charge density $\rho=\frac{q\delta(r-R)}{4\pi r^2}$. The electrostatic equation is now
\begin{equation}\label{33}
\frac{d}{dr}f_e\frac{d}{dr}A_0=q\delta(r-R)
\end{equation}
where $f_e=4\pi  e^{-\frac{\mu(r)}{2}} e^{-\frac{\nu(r)}{2}}r^2$. By taking $e^{\frac{\mu(r)}{2}} =e^{-\frac{\nu(r)}{2}}$, we find $f(r)=r^2$ (so unlike the scalar field, $A_0$ is only singular at $r=0$). The solution of (\ref{33}), salifying the boundary condition at $r_s$ is
\begin{equation}\label{34}
A_0= \left\{ \begin{array}{ll}
-\frac{q}{4\pi r}+\frac{q'}{4\pi r} ,& \textrm{$r\geq R$}\\
-\frac{q}{4\pi R}+\frac{q'}{4\pi r},& \textrm{$r_s\leq r\leq R$}\\
\end{array} \right.
\end{equation}
  where $q'=-\frac{r_s}{R}q$. This shows that the potential has two sources: the test charge $q$, and a charge $q'$ induced in the reflecting star. So in the presence of the external charge, the star has gained hair and adopted electric charge. By approaching $R$ to $r_s$, we have  $q'\to -q$, and for $R=r_s$  we find $A_0=0$ as a result of total zero net charge of the star. Hence we conclude that in the absence of any source, the asymptotically flat reflecting star has a null field in its background, while by adding some densities in the background it may adopt induced charge. The situation is the same for scalar and electric fields.

\subsection{$A(\phi)=M_0+\frac{\phi^2}{2M}$}

As a second example we choose $A(\phi)=M_0+\frac{\phi^2}{2M}$, where $M_0$ and $M$ are two real nonzero constants. $\frac{\rho}{M}$ behaves as an effective mass squared. The solution is now
\begin{equation}\label{35}
\phi(r)= \left\{ \begin{array}{ll}
\frac{\lambda} {M} \phi(R)\left(1-\frac{F(R)}{F(r_s)}\right) F(r),&  r\geq R\\
\frac{\lambda} {M} \phi(R)\left(1-\frac{F(R)}{F(r_s)}\right) F(R),&  r\leq R\\
\end{array} \right.
\end{equation}
For $R=r_s$, we have $\phi=0$, a sign of no-hair conjecture for a sole reflecting star without external matter. To have a non-trivial solution ($\phi(r)\neq 0$) it is necessary to have $R\neq r_s$ and
\begin{equation}\label{36}
\frac{\lambda}{M}  \left(1-\frac{F(R)}{F(r_s)}\right) F(R)=1.
\end{equation}
For the metric (\ref{27}), this gives
\begin{equation}\label{37}
\frac{\lambda}{16\pi M r_h}\ln\left(\frac{R-r_h}{R}\right)\left(1-\frac{\ln\left(\frac{R-r_h}{R}\right)}{\ln\left(\frac{r_s-r_h}{r_s}\right)}\right)=1.
\end{equation}
As $R>r_s>r_h$, we must have $\frac{\lambda}{M} <0$. This is in agrement with our general results in the second section and can be immediately obtained by setting $A_{\phi}=\frac{\phi}{M}$ and  $A_{,\phi\phi}=\frac{1}{M}$, in  (\ref{7}) or (\ref{10}).  Also by multiplying both sides of (\ref{12}) by $\phi$, and integrating from $r_s$ to infinity we obtain
\begin{eqnarray}\label{38}
-\int_{r_s}^\infty f(r)\left(\frac{d\phi}{dr}\right)^2 dr&=&\lambda \phi(R) A_{,\phi}(R)\nonumber \\
&=&\frac{\lambda}{M}\phi^2(R),
\end{eqnarray}
indicating that $\frac{\lambda}{M} <0$.

In the absence of the  surface $r_s$, the non-trivial solution, regular at the horizon $r_h<R$, is
\begin{equation}\label{39}
\phi(r) = \left\{ \begin{array}{ll}
\frac{\lambda}{M} \phi(R)F(r),&  r\geq R\\
\frac{\lambda}{M} \phi(R) F(R),&  r\leq R\\
\end{array} \right.
\end{equation}
This solution is regular on the horizon, as long as $R\neq r_h$. Again, the necessary condition to have this nontrivial solution is $\frac{\lambda}{M}F(R)=1$ which for (\ref{27}) becomes
\begin{equation}\label{40}
\frac{\lambda}{4\pi M r_h} \ln\left(1-\frac{r_h}{R}\right)=1.
\end{equation}
So a necessary condition to have scalar hair for the black hole (\ref{27}) is that the parameters must be chosen such that (\ref{40}) be satisfied. As an example let us tend the radius of the spherical charge to the horizon radius : $R\to r_h$. In this situation if we require that (\ref{40}) be satisfied and $\phi(r_h)$ be regular, then we must take  $\frac{\lambda}{M}\to 0$, leading to $\phi(r)\to 0$ for $r\geq r_h$ (see (\ref{39})).

\section{Conclusion}
The existence of scalar fields in the background of an asymptotically flat reflecting star was discussed. Reflecting star is a compact object, which instead of the absorbing event horizon, posses a boundary where the field vanishes.  Both the reflecting star and the black hole obey no scalar hair conjecture, but there is an essential difference between them: the reflecting star does not support the electric field either and has a null electric charge (see section 3.1).

Although black holes do not support scalar fields in their canonical minimally coupled form, this is not the case where the scalar field is non-minimally coupled to the geometry or to the matter surrounding a black hole. Is the same true for the reflecting star? This motivated us to consider a conformal coupling between the scalar field and the matter around the star, inspired by the previous works in the screening models (see (\ref{4}))\cite{screen1}. We derived some general results, which are very similar to the black hole case, showing that the presence of matter and the new coupling alters the basics of the Bekeinstein's theorem\cite{scalarmass} (see discussions after Eqs. (\ref{6})and(\ref{8})). Indeed by putting some condition on the coupling and matter density, one can evade no scalar hair conjecture.

Analytical solutions can be very enlightening in this regard, but unfortunately obtaining an analytical solution is not feasible for a realistic form of matter density. The use of a test scalar charge (see (\ref{11})), by allowing us to get an analytical answer, can be very promising. The solutions showed that the reflecting star behaves rather like a grounded black hole: in the presence of the conformally coupled scalar test charge, a continuous non-trivial scalar field appears (provided that the parameters satisfy (\ref{19})). This field consists of two parts: one originated from the test charge located outside the star, and the other one originated from an induced charge adopted by the star (see (\ref{17})).

 Note that the Dirichlet boundary condition implies that some matter (induced charge) inside the reflecting star acts as a source of the total scalar field vanishing on the surface. Otherwise, the scalar field originated only from the outside test charge would not be zero on the surface. This internal source does not appear in the local equation (\ref{5}) which is written for the outer space, but its role is derived from the boundary condition.

We showed also a black hole surrounded by the conformably coupled test charge admits a regular non-trivial scalar field, provided that the black hole and the test charge parameters satisfy some conditions (see (\ref{c})), although this does not mean that black hole adopts an additional primitive scalar charge.

\end{document}